\documentclass{amsart}

\usepackage[centertags]{amsmath}
\usepackage{amssymb}
\usepackage{amsthm}
\usepackage{amsfonts, latexsym, mathrsfs,array,color,booktabs}
\usepackage{graphicx}
\usepackage{float}

\usepackage{color}
\usepackage[noend]{algpseudocode}
\usepackage{algorithmicx,algorithm}

\begin{document}

\title[Blaschke Unwinding AFD and ECG compression]{A Novel Blaschke Unwinding Adaptive Fourier Decomposition based Signal Compression Algorithm with Application on ECG Signals}

\author{Chunyu Tan}
\address{Faculty of Science and Technology, University of Macau, Macao, China}
\email{yb57416@umac.mo}

\author{Liming Zhang}
\address{Faculty of Science and Technology, University of Macau, Macao, China}
\email{lmzhang@umac.mo}

\author{Hau-tieng Wu}
\address{Department of Mathematics and Department of Statistical Science, Duke University, Durham, NC, US}
\email{hauwu@math.duke.edu}

\maketitle

\begin{abstract}
This paper presents a novel signal compression algorithm based on the Blaschke unwinding adaptive Fourier decomposition (AFD). The Blaschke unwinding AFD is a newly developed signal decomposition theory. It utilizes the Nevanlinna factorization and the maximal selection principle in each decomposition step, and achieves a faster convergence rate with higher fidelity.
The proposed compression algorithm is applied to the electrocardiogram signal. To assess the performance of the proposed compression algorithm, in addition to the generic assessment criteria, we consider the less discussed criteria related to the clinical needs -- for the heart rate variability analysis purpose, how accurate the R peak information is preserved is evaluated.
The experiments are conducted on the MIT-BIH arrhythmia benchmark database. The results show that the proposed algorithm performs better than other state-of-the-art approaches. Meanwhile, it also well preserves the R peak information.
\end{abstract}

\section{Introduction}

The demand for efficient data transfer increases with the technological advances in mobile devices and its growing application in different fields, particularly the healthcare system.
Electrocardiogram (ECG) is ubiquitous in the health system, and it plays a key role in the mobile healthcare system. A distinguishing feature of the mobile health system is its potential to continuously record the ECG signal \cite{HDT, TTRDA}. A continuously recorded ECG signal allows a physician to find rare events that could not be easily found during a physician visit. Therefore, we would expect to record the signal as long as possible.
Due to the physical limitations, like the bandwidth and the battery support, a signal compression tool with a high compression quality is needed.

Adaptive Fourier decomposition (AFD) is a newly developed signal processing technique that generalizes the traditional Fourier decomposition, and there are several algorithms implementing AFD.
In addition to the Blaschke Unwinding AFD \cite{Q} selected in this paper, there are Core AFD \cite{QW}, the Cyclic AFD \cite{Q2}, etc. A detailed comparison of different AFD algorithms is presented in \cite{QLM}.
In general, unlike ordinary transform-based approaches based on a pre-selected basis, AFD decomposes a given analytic signal by adaptively choosing its associated basis from the Takenaka-Malmquist (TM) system according to some selection principle \cite{QW, QZL}. Due to its adaptivity to the signal, we may design an efficient signal compression tool with high reconstruction quality based on AFD.

The main contribution of this paper is introducing a novel transform-based signal compression algorithm based on the Blaschke unwinding AFD. Compared with other AFD algorithms, the Blaschke unwinding AFD particularly employs the Nevanlinna factorization to speed up the decomposition, and it has been theoretically validated for signal decomposition in the Hardy space $H^{2}$ and the Hilbert space $L^{2}$ \cite{Q, MDZQ}. The proposed algorithm is tested on a benchmark ECG database, the MIT-BIH arrhythmia database. The results show that it outperforms other existing state-of-the-art algorithms.
Another contribution of the paper is that we evaluate how well the compression algorithm approximates the original ECG signal for the purpose of heart rate variability (HRV) analysis. Particularly, we evaluate how accurate the R peak information is preserved. To the best of our knowledge, the relationship between a compression algorithm and HRV analysis is less discussed in the literature.

The paper is organized as follows. The related works are summarized in Section \ref{Section:RelatedWorks}. The proposed method is described in detail in Section \ref{Section:BUAFD}. In Section \ref{Section:ExpResults}, the proposed method is validated by  MIT-BIH arrhythmia benchmark database, and the comparisons with state-of-the-art algorithms are presented. Finally, conclusion is drawn in Section \ref{Section:Conclusion}.

\section{Related Works}\label{Section:RelatedWorks}

Compression methods are generally divided into two main categories \cite{AG,HWC}, lossy and lossless methods \cite{KA}. Lossy methods are more commonly used for practical signals due to their higher compression rate. Transform based approach is dominate in the lossy methods \cite{RM}. The principle is evaluating the coefficients of the original signal associated with a pre-designed basis so that the signal can be represented by few parameters. For example, the wavelet transform \cite{HWC,KKS,FF,TST}, the discrete cosine transform (DCT)  \cite{LKL,PSSS}, the Hermite transform \cite{DK,KH}, the nonlinear transform \cite{AP}, the compressed sensing \cite{Zhang2015,Polania2015,Craven2017}, and the singular value decomposition (SVD) \cite{Wei2001,Liu2017}.
While the wavelet transform is popular, however, the compression performance is largely dependent on the chosen mother wavelet, number of decomposition levels and different optimization schemes \cite{MD}. DCT is also frequently used because of its better energy compaction efficiency. However, the significant coefficients used for reconstruction are selected empirically without utilizing the intrinsic characteristic of the coefficients and the signal. The parameter optimizations of the Hermite transform rely on specific techniques that cannot guarantee the reconstruction fidelity.
Various types of nonlinear transforms suffer from the temporal and spatial complexities due to the application oriented preprocessing techniques, like an accurate QRS complex detection and segmentation. A similar limitation holds for the CS and the SVD approaches -- an establishment of the sensing matrix or a good beat detection is needed as a pre-processing step.
While these techniques have been widely applied and have their own merits, an algorithm that unifies their benefits, like being adaptive to the signal, free of any application oriented preprocessing technique, and efficient, is called for.

AFD-based approach could satisfy these requirements. However, to the best of our knowledge, there is only one paper taking the AFD into account \cite{MZD}. It is adaptive to the signal, free of any application oriented preprocessing technique, and reported to outperform the state-of-the-art algorithms \cite{MZD}. Compared with our algorithm, the authors in \cite{MZD} count the Core AFD and combine it with a symbol substitution technique to compress the ECG signal.
Since it has been shown that the Blaschke unwinding AFD method is more effective than the Core AFD, both theoretically and numerically \cite{QLM}, we claim that the Blaschke unwinding AFD could lead to a better result.
A comparison of the Core AFD and the Blaschke unwinding AFD can be found in Section \ref{Comparison:Algo} below.

\section{Blaschke unwinding AFD based Compression Algorithm}\label{Section:BUAFD}

We start from a high level overview of the algorithm. The Blaschke unwinding AFD is a novel signal processing technique developed in the last decade \cite{Q}, which is a nonlinear generalization of the Fourier series. To illustrate how the generalization works, take a $L^1$ function $f$ defined on $[0,1]$ with $\hat{f}(-k)=0$ for all $k\in\mathbb{N}$ as an example. $f$ could be viewed as a function defined on $\partial\mathbb{D}=\{z\in \mathbb{C}|\,\|z\|=1\}$ coming from an analytic function $F_0(z)$ defined on the open unit disk $\mathbb{D}:=\{z\in\mathbb{C}|\,\|z\|<1\}$ with the relation $f(\theta)=\lim_{r\to 1}F_0(re^{i2\pi\theta})$. The Fourier series of $f$ could be viewed as peeling off $0$ as a root from $F$ step by step. Indeed, we have
\begin{align}
F_0(z)&=F_0(0)+(F_0(z)-F_0(0))=F_0(0)+zF_1(z)\nonumber\\
&=F_0(0)+zF_1(0)+z(F_1(z)-F_1(0))=\ldots\nonumber\\
&=\sum_{l=0}^nz^lF_l(0)+z^n(F_n(z)-F_n(0))\,,\label{Motivation:FT}
\end{align}
where the second equality comes from the fact that $0$ is a root of $F(z)-F(0)$ and $F_l(z):=\frac{F_{l-1}(z)-F_{l-1}(0)}{z}$ for $l=1,2,\ldots$, and by restricting $F$ to $\partial\mathbb{D}$,
\begin{align}\label{Motivation:FT1}
f(\theta)=F(e^{i2\pi\theta})= \sum_{l=0}^\infty e^{i2\pi l\theta}F_l(0)\,,
\end{align}
we get the Fourier series. Thus, from the complex analysis viewpoint, the Fourier series comes from iteratively peeling off the root $0$ from the signal. However, $0$ might not be the only root of $F_1$. It thus raises a natural question -- what if we take roots other than $0$ into account to decompose the signal? For example, for $F_1$, if we take some root $a_1$ so that some good conditions hold and iterate the process, could we achieve a more adaptive decomposition of the signal with a faster convergence rate? Or even more generally, could we find a canonical decomposition of a given analytic function based on the roots of the function? This question actually has a positive answer, which is widely known as the Nevanlinna factorization \cite{G}.
Based on the Nevanlinna factorization, the above-mentioned idea led to several theoretical and algorithmic development in the past decade for the signal processing purpose \cite{Nahon-2000,Q,CS,CSW,MDZQ}, and the Blaschke unwinding AFD is a specific example.
In short, based on the Nevanlinna factorization of Hardy space functions \cite{G} the Blaschke unwinding AFD is composed of iterating the following three steps. First, we find an orthogonal system, the TM system \cite{Takenaka,Malmquist}, to describe possible roots; second, we follow a principle, called the {\em maximal selection principle}, to select a suitable root; third, we decompose the signal into two parts. The resulting decomposition that generalizes Fourier series is shown in Equation (\ref{004}) below.

\subsection{Mathematical Foundation}

\begin{figure}[!htbp]
  \centering
  \includegraphics[width=.99\linewidth]{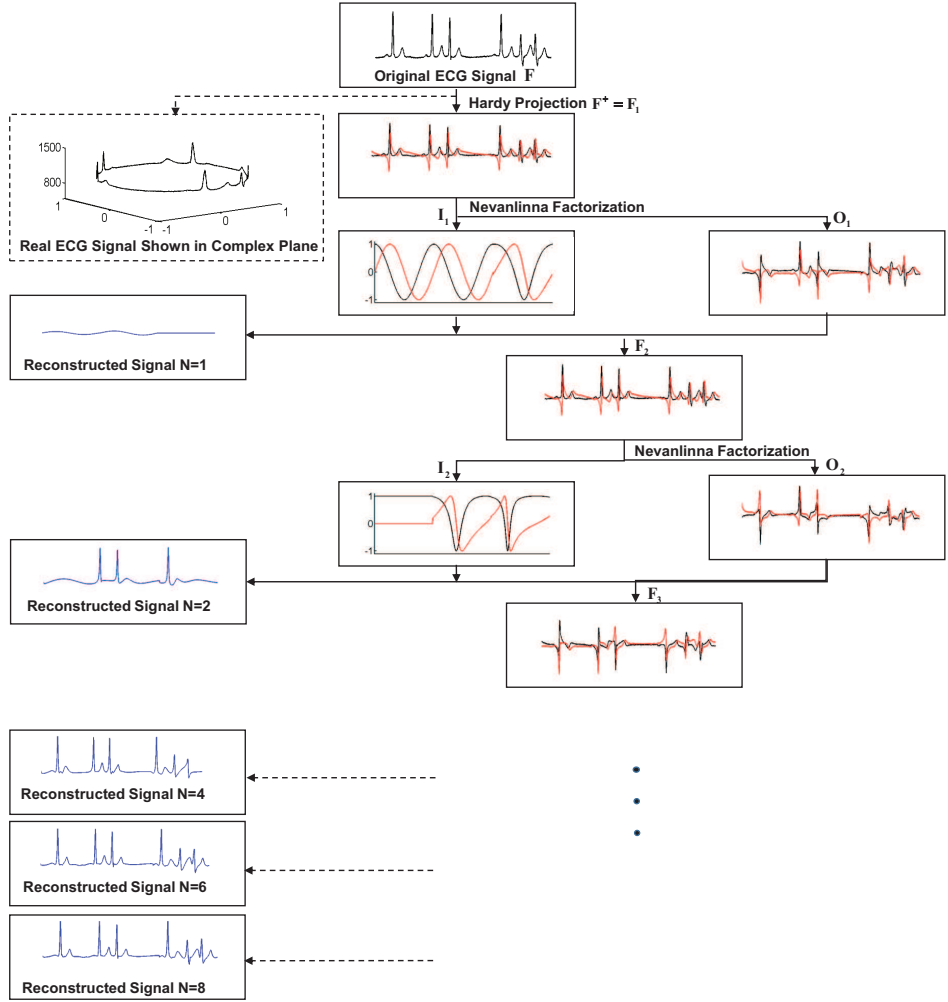}
  \caption{The flowchart of the Blaschke unwinding AFD with a real ECG signal. The black and red curves denote the real and imaginary parts of a complex signal, respectively. Reconstruction signals with decomposition level $N=2,4,6,8$ are shown on the left hand side to illustrate the fast convergence of the algorithm.}
  \label{fig:flowchart}
\end{figure}

\begin{figure}[!htbp]
  \centering
  \includegraphics[width=.99\linewidth]{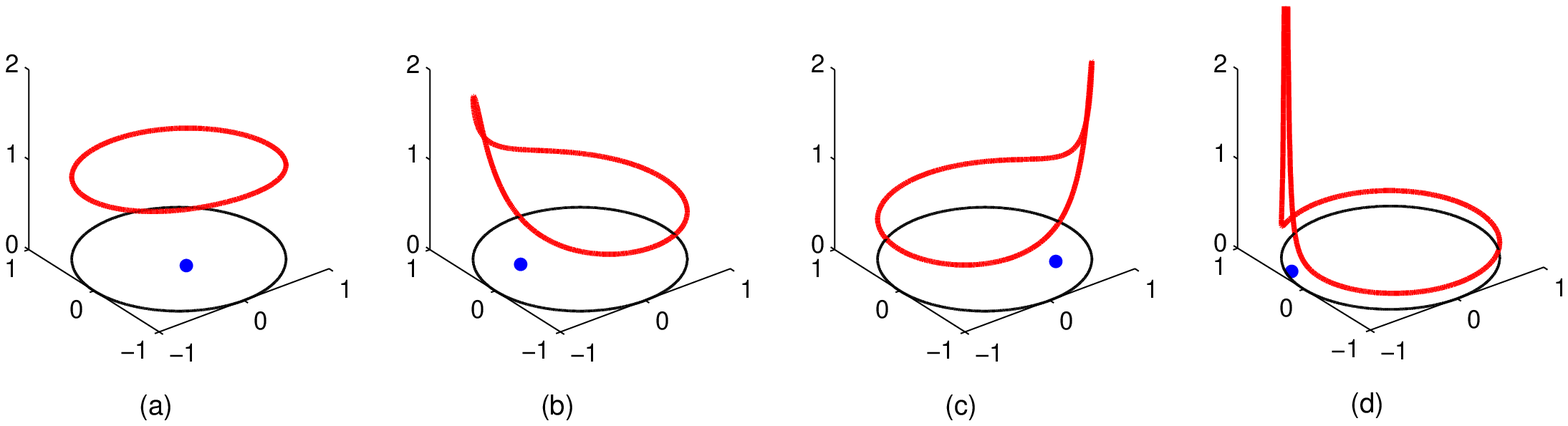}\\
  \caption{Visualization of $e_a$ in the complex plane with different $a$; (a) $a=-0.02-0.14i$, (b) $a=-0.5+0.26i$, (c) $a=0.5-0.44i$, (d) $a=-0.88+0.34i$. The red curve denotes the real part of $e_{a}(e^{it})$, where $t\in [0,2\pi)$; that is, the restriction of $e_a$ on the unit circle. The black curve denotes the unit circle, and the dot inside the disk denotes the associated $a$.}
  \label{fig:picture01}
\end{figure}

In this section, we summarize the mathematical foundation of Blaschke unwinding AFD. The algorithm is illustrated with a real ECG signal in Figure \ref{fig:flowchart}. Readers with interest in the algorithm could jump to Section \ref{Section:AlgorithmStep}.
For a given set of $a_{n}\in \mathbb{D}$, where $n=1,2,\ldots$, the rational orthogonal system, or the TM system, is defined as $\{B_{n}\}$, where
\begin{equation}\label{000}
  B_{n}(z)=\frac{\sqrt{1-|a_{n}|^{2}}}{1-\bar{a}_{n}z}\prod_{k=1}^{n-1}\frac{z-a_{k}}{1-\bar{a}_{k}z},
\end{equation}
and $B_{n}$ are called \em modified Blaschke products\em.
We call a TM system {\em adaptive} if we select $a_{n}$ according to the input signal.
An adaptive TM system is a generalization of the {\em Fourier system}, which is composed of the set of polynomials $\{z^{n}\}_{n=0}^{\infty}$. Note that it corresponds to $a_{n}=0$ for all $n$ in the TM system.
We call an element in the dictionary $\{e_{a}\}_{a\in\mathbb{D}}$, where
\begin{equation}\label{0000}
 e_a =\frac{\sqrt{1-|a|^{2}}}{1-\bar{a}z},
\end{equation}
an {\em evaluator}, which is  employed to facilitate computation of energy gain during the decomposition. See Figure \ref{fig:picture01} for examples of $e_a$ with different $a$'s in the complex plane. Clearly, a modified Blaschke product in (\ref{000}) comes from a product of several evaluators with a proper normalization. The evaluators are also the reproducing kernels of the Hardy space $H^{2}$.

Take an analytic function $F^+\in H^2$. By the Nevanlinna Factorization Theorem \cite{G}, $F^{+}$ can be decomposed into an outer function and an inner function so that
\begin{equation}\label{a5}
F^{+}=IO\,,
\end{equation}
where $O(z):=\exp\{ \frac{1}{2\pi}\int_0^{2\pi}\frac{e^{it}+z}{e^{it}-z}\log|F^+(e^{it})|dt\}$ is the outer function and $I(z):=F^+(z)/O(z)$ is the inner function. Clearly, the outer function is non-zero in $\mathbb{D}$. For the practical purpose, we reasonably assume that the singular part of the inner function equals $1$. As a result, the inner function contains all roots of $F^+$ in $\mathbb{D}$, which is actually the (possibly infinite) Blaschke product formed using the zeros of $F^+$ in $\mathbb{D}$. In other words, the function $F^+$ is decomposed into the product of an outer function and a Blaschke product. This decomposition is unique up to a constant with norm $1$. Although not used in this paper, we mention that the inner function $I$ has a non-negative analytic instantaneous frequency \cite[Theorem 1.1]{Nahon-2000,Q}.

The key step in the Blaschke unwinding AFD is decomposing the analytic function $F^{+}$ into the form
\begin{equation}\label{002}
  F^{+}(z)=\sum_{n=1}^{\infty}c_{n}I_{(n)}B_{n}\,,
\end{equation}
by taking the Nevanlinna factorization into account, where $c_{n}$ are constants, $I_{(n)}$ are inner functions and $B_{n}$ are modified Blaschke products as defined in (\ref{000}).
To achieve (\ref{002}), we start from $F^{+}=F_{1}=I_{1}O_{1}$.
To further decompose $O_{1}$ with a fast convergence rate, we apply the reproducing kernel property of the evaluator and the {\em maximal selection principle}.
By the reproducing kernel property of the evaluator, the inner product between $O_{1}$ and $e_{a}$ satisfies \cite{QW}
\begin{equation}\label{a6}
  \left\langle O_{1},e_{a}\right\rangle=\sqrt{1-|a|^{2}}O_{1}(a).
\end{equation}
The maximal selection principle proved in \cite{QW} asserts that for $O_{1}$, there exists $a_{1}\in \mathbb{D}$ such that
\begin{equation}\label{003}
  a_{1}=\arg\max_{a\in \mathbb{D}}\{(1-|a|^{2})|O_{1}(a)|^{2}\}.
\end{equation}
Therefore, we can further decompose $F_1$ by \cite{Q}
\begin{equation}\label{a7}
  F_{1}(z)=I_{1}(z)\left\langle O_{1},e_{a_{1}}\right\rangle B_{1}(z)+I_{1}(z)\frac{z-a_{1}}{1-\bar{a}_{1}z}F_{2}(z)\,,
\end{equation}
where $F_2(z):=\frac{O_{1}-\left\langle O_{1},e_{a_{1}}\right\rangle e_{a_{1}}}{\frac{z-a_{1}}{1-\bar{a}_{1}z}}$.
We note that the reminder $F_{2}(z)$ is still in $H^{2}$. We emphasize that in the Fourier series (\ref{Motivation:FT}), we consider the root $0$ instead of $a_1$.
Repeating the same process, we obtain, after $N$ steps,
\begin{equation}\label{004}
  F^{+}(z)=\sum_{n=1}^{N}c_{n}I_{(n)}(z)B_{n}(z)+R_{N}(z)\,,
\end{equation}
where $c_{n}=\left\langle O_{n},e_{a_{n}}\right\rangle$, $I_{(n)}=\prod_{i=1}^{n}I_{i}$, $R_{N}$ is the remainder after $N$ times decomposition, and $a_{n}\in \mathbb{D}$ is selected according the maximal selection principle (\ref{003}).

Denote
\begin{equation}\label{0004}
F^{+}_{rec}=\sum_{n=1}^{N}c_{n}I_{(n)}B_{n}.
\end{equation}
We call $N\in \mathbb{N}$ {\em decomposition level}. It has proved in \cite[Theorem 2.2]{Q} that $F^{+}_{rec}$ converges to $F^{+}$ in the $H^{2}$ convergence sense; that is, $\|R_{N}\|_{H^{2}}\rightarrow0$ as $N\rightarrow\infty$.
Therefore, according to (\ref{001}), $F_{rec}$ approximates the original real-valued signal $F$. Although the convergence rate of the approximation is still open, we found that in experiments, the error term decays according to the exponential rule; that is, $\|R_{N}\|_{H^{2}}=O(\exp (-\delta N))$ for some $\delta >0$ \cite{CS}. If the associated analytic signal can be analytically extended to outside of the closed unit disc, then this high decay rate can be indeed shown rigorously \cite{Q}. This property of unwinding decomposition defensively makes the high efficiency of its application to signal compression.
Last but not least, the Blaschke-unwinding method can effectively denoise the signal due to the integration effect and robust nature of complex analytic functions.

In summary, the Blaschke unwinding AFD generates an adaptive basis by pursuing the maximal energy gain of the outer function in each of the decomposition iteration. Although theoretically it is an open and difficult problem, numerically it ``consistently'' converges fast and robust in both the energy and the pointwise measurements. Benefited from these characteristics, the Blaschke unwinding AFD has a potential for decomposing multifarious signals including ECG signals as well as others for the e-health oriented applications.

\begin{algorithm}[t]
\caption{Blaschke Unwinding AFD based Compression}
{\bf Input:}
 Real-valued input signal $F$, sets of parameters $a\in\mathbb{D}$ and the decomposition level $N$.\\
{\bf Output:}
Parameters $\{c_{n}\}_{n=1}^{N}$, $\{a_{n}\}_{n=1}^{N}$ and a finite number of zeros $\{\{r_{n_{j}}\}_{j=1}^{M_{n}}\}_{n=1}^{N}$.
\begin{algorithmic}[1]
\State Get the projection signal $F^{+}$ of $F$ in the Hardy space.
\State Initialize $F_{1}=F^{+}$.
\For{$n=1$ to $N$} 
\State Obtain the inner function  $I_{n}$ and outer function $O_{n}$ of $F_{n}$ so that $F_{n}=I_{n}O_{n}$;
\State Get zeros $\{r_{n_{j}}\}_{j=1}^{M_{n}}$ of $I_{n}$ by Algorithm \ref{Algorithm2};
\State Get $a_{n}=\arg\max\{(1-|a|^{2})|O_{n}(a)|^{2}:\,a\in \mathbb{D}\}$;
\State Get $c_{n}=\left\langle O_{n},e_{a_{n}}\right\rangle$;
\State Get $F_{n+1}=\frac{F_{n}-c_n I_{n} e_{a_{n}}}{I_{n}}\frac{1-\bar{a}_{n}z}{z-a_{n}}$;
\EndFor
\State \Return $\{c_{n}\}_{n=1}^{N}$, $\{a_{n}\}_{n=1}^{N}$, $\{\{r_{n_{j}}\}_{j=1}^{M_{n}}\}_{n=1}^{N}$.
\end{algorithmic}
\end{algorithm}

\begin{algorithm}[t]
\caption{Procedure for calculating zeros\label{Algorithm2}}
{\bf Input:} $F$, $\delta>0$\\
{\bf Output:} zeros of $F$, $\{r_{j}\}_{j=1}^{M}$.
\begin{algorithmic}[1]
\State Determine $M$ by $M=\frac{1}{2\pi i}\int_{|z|=1}\frac{F'(z)}{F(z)}dz$.
\State  Initialize $G_{1}:=F$.

\For{$j=1$ to $M_{1}$}

\State Evaluate $\arg\min_{z\in\mathbb{D}_{1-\delta}}|G_{j}(z)|$ ;
\State Get $r_{j}$ satisfying $G_{j}(r_{j})=0$ ;
\State Get $G_{j+1}:=G_{j}\frac{1-\bar{r}_{j}z}{z-r_{j}}$ ;

\EndFor

\State \Return $\{r_j\}_{j=1}^{M}$.
\end{algorithmic}
\end{algorithm}

\subsection{Proposed Compression Algorithm}\label{Section:AlgorithmStep}
The compression consists of two steps. The first step carries out the Hardy projection and the Blaschke unwinding AFD compression. The second step is the lossless Huffman encoding. For the decompression, it is the inverse of the compression, including the Huffman decoding and the inverse Blaschke unwinding AFD process.

\subsubsection{Implement the Blaschke unwinding AFD}

Algorithm 1 illustrates how the Blaschke unwinding AFD is applied to compress a real-valued signal.
First, the input real-valued signal $F$ is projected to $H^{2}$ space and we get $F^{+}$.
In practice, we could safely assume that
\begin{equation}\label{001}
2\Re F^{+}=F+c_{0}
\end{equation}
holds, where $\Re$ means taking the real part, and $c_{0}$ is the zero-th Fourier coefficient of $F$. $c_0$ is the first data point we save for the signal compression, and
$F^{+}$ is initialized  as the first remainder $F_{1}$.
Second, extract the inner function by calculating zeros of $F_{1}$ by the method introduced in \cite{MDZQ}, where we assume that $F_{1}$ has finite roots on $\overline{\mathbb{D}}:=\{z\in\mathbb{C}|\,\|z\|\leq1\}$ \cite{MDZQ}. The detailed steps of numerical calculation for calculating zeros of $F_{1}$  are performed in Algorithm 2.
Then accordingly, get the outer function $O_{1}$ by the Nevanlinna factorization.
Third, The set of $\{a_{n}\}$, $n=1,2,\ldots,$ in (\ref{000}) consisting of discrete points in $\mathbb{D}$ is generated by dividing $\mathbb{D}$ into rectangular grid to get the TM system in (\ref{000}) and evaluators $\{e_{a}\}$ in (\ref{0000}). Then, the decomposition of $O_{1}$ is based on the TM system. During the decomposition, the maximal selection principle is applied in the selection of $a_{1}$ with the aid of evaluators.
Suppose the decomposition level is $N\in\mathbb{N}$. Iterate the above three steps, each on the remainder of the previous step, for $N$ times, and we end up with $\{e_{a_n}\}$, the modified Blaschke products $\{B_{n}\}$, and $M_n$ zeros, for $n=1,\ldots,N$.
As a result, we obtain $2N+1$ parameters, including $\{c_{n}\}_{n=0}^{N}$, and $\{a_{n}\}_{n=1}^{N}$, as well as $\sum_{n=1}^{N}M_{n}$ zeros. $c_n$ and $a_n$, where $n=1,\ldots,N$, as well as $\sum_{n=1}^{N}M_{n}$ zeros, are other data points we save for the data compression.

\subsubsection{Quantization and Lossless Huffman Encoding}
The quantization step is carried out for those $2N+1$ parameters, $\{c_{n}\}_{n=0}^{N}$, $\{a_{n}\}_{n=1}^{N}$, and $\sum_{n=1}^{N}M_{n}$ zeros for the lossless encoding. It includes extracting a small threshold to get rid of parameters that are close to zero, and quantifying all parameters to have the same order of magnitude. For the ECG signal, we choose to achieve it by
\begin{align}\label{n1}
  \bar c_{n}:=&\texttt{Round}(c_{n}), \nonumber\\
  \bar a_{n}:=\texttt{Round}(100\times a_{n}), &\bar r_{n_{j}}:=\texttt{Round}(100\times r_{n_{j}}),
  \end{align}
where $\texttt{Round}()$ denotes rounding numbers, $n=1,\ldots, N$, and $j=1,\ldots,M_n$.
The quantized data, $\{\bar c_{n}\}_{n=0}^{N}$, $\{\bar a_{n}\}_{n=1}^{N}$ and $\{\{\bar r_{n_{j}}\}_{j=1}^{M_{n}}\}_{n=1}^{N}$, are coded by the Huffman coding algorithm.

\subsubsection{Reconstruction of the Compressed Data}

In the decompression stage, the reconstruction of the compressed data is accomplished by inverting the two steps in the compression stage. The first step is applying the Huffman decoding, which refers to the defined data table. The second step is to use the compression items being weighted summed and back projected to get the reconstructed real-valued signal; that is, we have
\begin{equation}\label{b1}
 F_{rec}^{+}(z):=\sum_{n=1}^{N}\bar c_{n}\bar I_{(n)}\bar B_{n}\,,
\end{equation}
where $\bar B_n$ is the modified Blaschke product constructed from $\{\bar a_{n}\}_{n=1}^{N}$ and $\bar I_{(n)}$ is the inner function constructed from $\{\{\bar r_{n_{j}}\}_{j=1}^{M_{n}}\}_{n=1}^{N}$.
The reconstructed real-valued signal is
\begin{equation}\label{b2}
  F_{rec}=2\Re F_{rec}^{+}-\bar c_{0}=2\Re \sum_{n=1}^{N}\bar c_{n}\bar I_{(n)}\bar B_{n}-\bar c_{0}.
\end{equation}

\subsubsection{Computational Complexity of the Proposed Algorithm}

The complexity of the proposed algorithm is as follows. Suppose the input ECG signal is of length $L\in\mathbb{N}$.
\begin{itemize}
\item The computational complexity of the Hardy projection is $\mathcal{O}(L)$.
\item To get the formula (\ref{a6}) and (\ref{003}), the computational complexity is $\mathcal{O}(ML^{2})$ as is shown in \cite{QZL}, where $M$ is the number of discrete points in $\mathbb{D}$ that is divided into rectangular grid for searching the maximal value to satisfy the formula (\ref{a6}).
\item The computational complexity of calculating the number of zeros, $M_{n}$, in Algorithm 1 is $\mathcal{O}(L)$. For the approximation formula of step 4 in Algorithm 2 to find zeros, we need to divide $\mathbb{D}$ into rectangular grid for searching the minimal value as approximate values of zeros. For the sake of simplicity,  taking the $M$ points same as the previous step and the computational complexity is $\mathcal{O}(M)$.
\end{itemize}
Therefore, the computational complexity of the Blaschke unwinding AFD is $\mathcal{O}(L^{2})$ since $M$ is determined. Note that the computational complexity is same with the core AFD and the discrete Fourier transform (DFT) \cite{QZL} and is expected to reduce to $\mathcal{O}(L\log L)$ as in \cite{GKQW} that incorporates fast Fourier transform (FFT) in the AFD algorithm. Moreover, as reported in \cite{LLW}, the computational complexity of the discrete wavelet transform (DWT) based compression algorithm and the CS based compression algorithm are $\mathcal{O}(L\log L)$ and $\mathcal{O}(L)$ respectively. The Nevanlinna factorization step gives rise to high computation complexity of the Blaschke unwinding AFD, however, is compensated by fast convergence.

\subsection{Comparison of Nevanlinna factorization based algorithms}\label{Comparison:Algo}

Based on the Nevanlinna factorization, there have been several algorithms available in the field. In this section, we provide a comparison of the Blaschke unwinding AFD and other Blaschke decomposition algorithms, particularly the Core AFD \cite{QZL,MZD} and the Blaschke decomposition algoritm \cite{CS,Nahon-2000,CSW}.

At first glance, the Core AFD method \cite{QZL} considered in the compression algorithm proposed in \cite{MZD} is close to the Blaschke unwinding AFD. However, there is a fundamental difference. In Core AFD, the Nevanlinna factorization is not considered for the decomposition. The decomposition only depends on the maximal selection principle. The Blaschke unwinding method \cite{Q,CS}, on the other hand, achieves a faster convergence in signal representation by eliminating inner function part in each iteration based on the Nevanlinna factorization.

The Blaschke decomposition algorithm considered in \cite{CS,Nahon-2000,CSW} is directly related to the Blaschke unwinding AFD in the sense that the Nevanlinna factorization is considered.
However, in the Blaschke decomposition algorithm, the TM system is not directly applied (zeros are not found); instead, in the $n$-th iteration, the outer function $O_n$ of $F_n$ in Algorithm 1 is estimated, the mean of $O_n$ is considered to produce zeros inside the unit disk, and the inner function is obtained through the division $I_n:=F_n/O_n$.
In \cite{CS,Nahon-2000,CSW,Saito}, the outer function is computed by the Hilbert transform of the logarithm of the absolute value of the analytic signal, viz., $e^{H(\ln |F|)}$, where $H$ stands for the Hilbert transformation operator.\footnote{To treat the computational unstabilization of the Hilbert transform when some values of $F$ being close to zero, the thesis \cite{Nahon-2000} proposes to use the technique involving a positive sequence $\epsilon\to 0+,$ viz.,
$H(\ln |F|)=\lim_{\epsilon\to 0+}H(\ln (|F|+\epsilon))$,
whereas Letelier and Saito proposed adding a small pure sinusoid \cite{Saito}.}
Compared with the Blaschke decomposition algorithm, instead of approximating the outer function and hence the inner function, the Blaschke unwinding AFD directly computes roots to approximate the inner function by a finite Blaschke products \cite{MDZQ}.
Note that with roots of the inner function at each iteration and the fast convergence property of the Blaschke unwinding AFD, the original signal can be efficiently compressed into a few coefficients and zeros. While this compression property is based on finding roots by the TM system, it is not transparent how to compress the signal directly from the output of the Blaschke decomposition algorithm.

\section{Experiment Results}\label{Section:ExpResults}

\subsection{Experiment Settings}

The ECG database employed in our experiments is the widely used MIT-BIH arrhythmia database \cite{MIT}. The database consists of 48 patient records, each thirty minutes in lengths. The recordings are sampled at the frequency of 360Hz with the resolution of 11 bits.
While there are two leads signal in the database, we follow the convention of most reported ECG compression methods and focus on the first lead.
Each record is separated into contiguous and non-overlapping windows for compression. For avoid long latency for real applications, we choose a shorter window of $600$ samples in our simulation. Unlike the segmentation in most transform-based compression methods, we do not need any ECG preprocessing, such as the beat detection. The experiments are conducted on a computer with 16GB RAM and 2.71 GHz Inter Core i5 processor and the code is implemented in MATLAB.

\begin{table}[!h]
\centering\caption{Performance of the proposed compression algorithm evaluated on 48 ECG Records with different decomposition level $N$. \label{Table:PerformanceDifferentN}}
\scalebox{0.9}{
\begin{tabular}{lcccccccc}
\toprule
N            & 7     & 8     & 9      &10     & 11     & 12  &13  &14  \\\midrule

CR           &42.27 & 35.53 & 30.21  & 25.99  & 22.80  & 20.38 &18.48 &16.85\\
PRD(\%)      &1.71   &1.47   &1.29    &1.14    &1.03    &0.94&0.87&0.81\\
QS           &33.41  &32.58  &31.53  &30.23   &29.19   &28.44 &27.84&27.28\\
SNR          &28.08 &31.05   &33.80   &36.08   &38.12   &39.87  &41.48 &43.02\\
Comp.time(s) &0.7857 &0.8720 &0.9724 & 1.0554 & 1.1462 &  1.2435 & 1.3379  &1.4320\\
Decomp.time(s) &0.0013& 0.0014& 0.0016& 0.0017 & 0.0019 & 0.0022& 0.0025  & 0.0026\\\bottomrule
\end{tabular}}
\end{table}

\begin{figure}[!h]
  \centering
  \includegraphics[width=.7\linewidth]{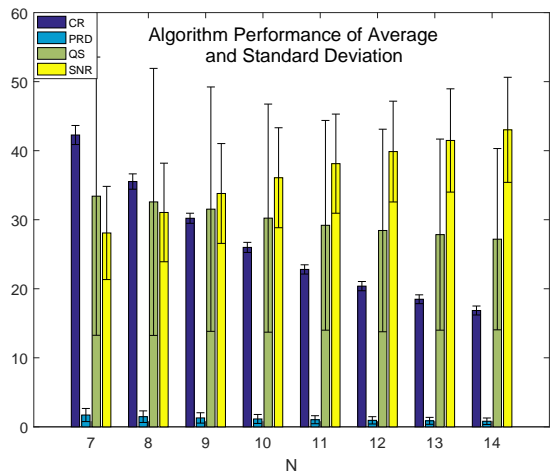}
  \caption{Error bar plot of four performance metrics of the proposed compression algorithm evaluated on 48 ECG records with different decomposition level $N$.}
  \label{fig:picture000}
\end{figure}

\begin{table*}[!htbp]
\centering\caption{Performance of the proposed compression algorithm evaluated on 48 ECG records  with $N=8$.\label{Table:PerformanceN=8}}
\scalebox{0.9}{
\begin{tabular}{ccc|ccc|ccc|ccc}
\toprule
Data   & CR         & PRD(\%)      &Data   &CR          & PRD(\%)        &Data    &CR          & PRD(\%)     &Data    &CR          & PRD(\%)   \\\midrule
100    & 39.34      &0.71          &114    & 36.92      &0.48            &203     &36.92       &2.09         &221     &35.29       &1.02 \\
101    & 36.64      &0.89          &115    & 35.29      &1.50            &205     &36.92       &0.70         &222     &36.09       &0.70 \\
102    & 36.09      &0.74          &116    & 35.29      &2.93            &207     &35.29       &0.63         &223     &35.29       &1.30 \\
103    & 35.82      &1.37          &117    & 34.78      &1.57            &208     &34.78       &1.49         &228     &36.64       &0.96 \\
104    & 34.78      &1.06          &118    & 35.29      &2.55            &209     &34.78       &1.44         &230     &34.28       &2.22 \\
105    & 35.82      &1.45          &119    & 34.78      &2.37            &210     &36.36       &0.88         &231     &35.29       &1.87 \\
106    & 34.78      &1.59          &121    & 35.82      &0.77            &212     &34.29       &1.91         &232     &35.56       &1.07 \\
107    & 34.29      &3.59          &122    & 36.09      &2.36            &213     &34.78       &4.34         &233     &35.29       &3.22 \\
108    & 35.82      &0.54          &123    & 34.29      &1.46            &214     &35.82       &0.86         &234     &36.36       &1.52  \\
109    & 35.04      &2.21          &124    & 34.78      &0.98            &215     &35.82       &1.92         &average &35.53       &1.47  \\
111    & 37.80      &0.78          &200    & 33.80      &0.87            &217     &34.29       &1.64         &std.    &1.11        &0.84 \\
112    & 36.09      &0.93          &201    & 37.80      &0.68            &219     &35.29       &1.38          \\
113    & 34.29      &1.56          &202    & 33.57      &0.31            &220     &34.78       &1.45         \\\bottomrule
\end{tabular}}
\end{table*}

\subsection{Performance Evaluation Metrics}

We consider the following measurements to evaluate of the proposed compression algorithm \cite{LKL} -- the compression ratio (CR), the percentage root-mean-square difference (PRD), the quality score (QS) and the signal to noise ratio (SNR).

CR measures the ratio between the bits of the original signal, $N_{inp}$, and the bits of the reconstructed signal, $N_{out}$, which is defined as
\begin{equation}\label{1}
  CR=\frac{N_{inp}}{N_{out}}.
\end{equation}
A larger CR indicates the higher compression efficiency.
PRD measures the difference between the original signal $X_{s}(n)$ and the reconstructed signal $X_{r}(n)$, which is defined as
\begin{equation}\label{2}
  PRD(\%)=100\times\sqrt{\frac{\sum_{n=1}^{N}(X_{s}(n)-X_{r}(n))^{2}}{\sum_{n=1}^{N}(X_{s}(n))^{2}}}.
\end{equation}
QS reflects the tradeoff between the compression efficiency and the reconstruction quality, which is defined as:
 \begin{equation}\label{6}
   QS=\frac{CR}{PRD}.
 \end{equation}
QS quantifies the overall compression performance \cite{FG}.
SNR is defined as
\begin{equation}\label{5}
  SNR=10\times\log\Big(\frac{\sum_{n=1}^{N}(X_{s}(n)-\bar{X})^{2}}{\sum_{n=1}^{N}(X_{s}(n)-X_{r}(n))^{2}}\Big).
\end{equation}
Where $N$ is the total number of data instances and $\bar{X}$ is the mean value.

The ECG morphology is the main focus when a physician conducts his clinical diagnosis. We focus on the QRS waveform reconstruction to further evaluate how accurate our algorithm preserves the clinical information inside the ECG signal, like the R-peak to R-peak interval. To achieve this goal, we evaluate how the compression procedure influences the commonly applied R peak algorithm. We consider the following metrics, including sensitivity (Se), positive predictive value (PPV) and $F_{1}$-measure. They are considered to evaluate how well the R peaks are preserved \cite{Standard2008}.
Se reports the ability to correctly detect the annotated QRS complexes, which is defined as
\begin{equation}\label{06121}
  Se=100\times\frac{TP}{TP+FN}.
\end{equation}
where TP and FN are the true positive (correctly identified QRS) and the false negative (missed QRS) detections.
PPV measures the ability to correctively identify the non-QRS complex, which is defined as
\begin{equation}\label{06122}
  PPV=100\times\frac{TP}{TP+FP}.
\end{equation}
where FP is the false positive (extra falsely detected QRS).
For the algorithm parameter optimization, an $F_{1}$-measure is used as suggested in \cite{BJJ}, which is defined as
\begin{equation}\label{06123}
  F_{1}=100\times\frac{2PPV\cdot Se}{PPV+Se}=100\times\frac{2TP}{2TP+FN+FP}.
\end{equation}
The $F_{1}$ statistic is an harmonic mean of the Se and PPV, which provides a more realistic measure of the performance.

In summary, an efficient signal compression tool with high reconstruction quality for ECG signals is expected to have a larger CR, QS, and SNR value, and smaller PRD value, besides, larger Se, PPV and $F_{1}$ value reflect higher QRS detection accuracy.

\begin{figure}[!t]
  \centering
  \includegraphics[width=.8\linewidth]{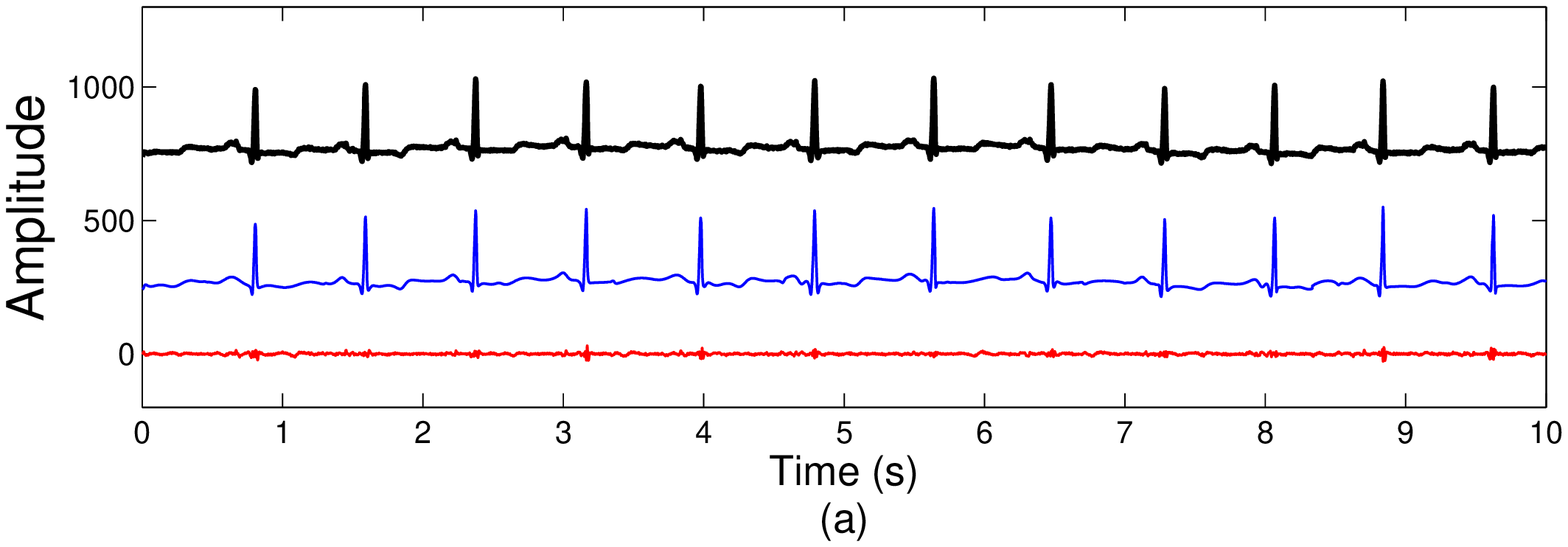}
  \includegraphics[width=.8\linewidth]{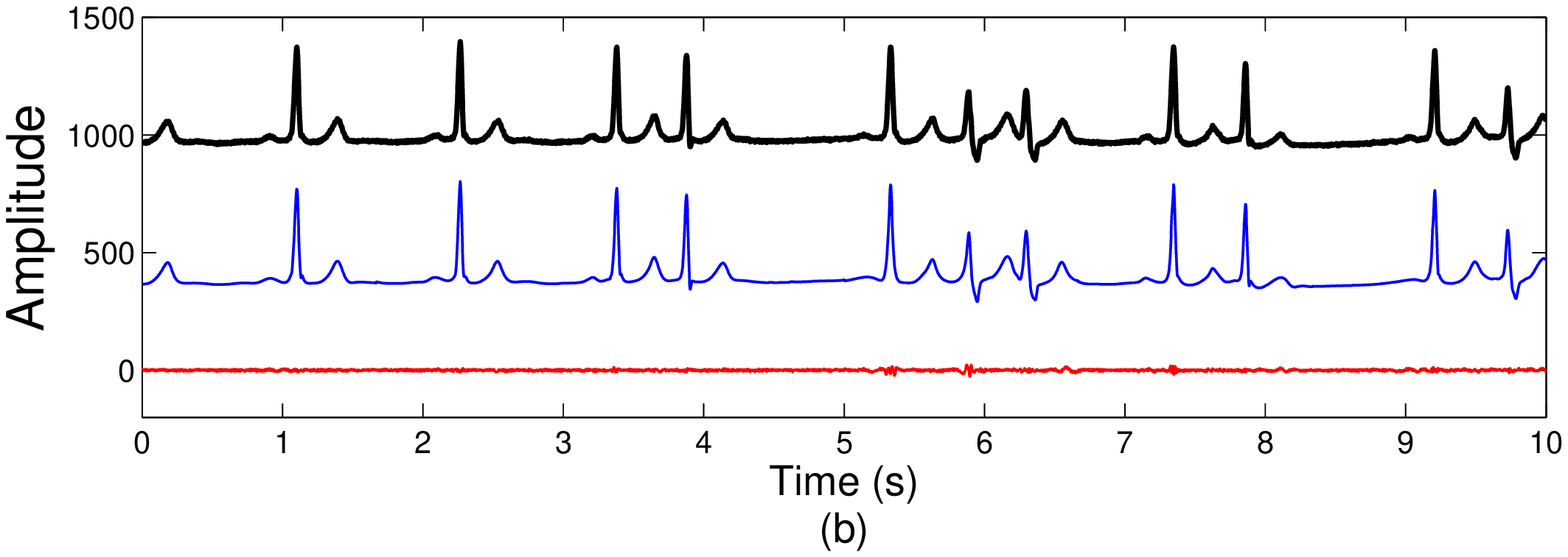}
  \includegraphics[width=.8\linewidth]{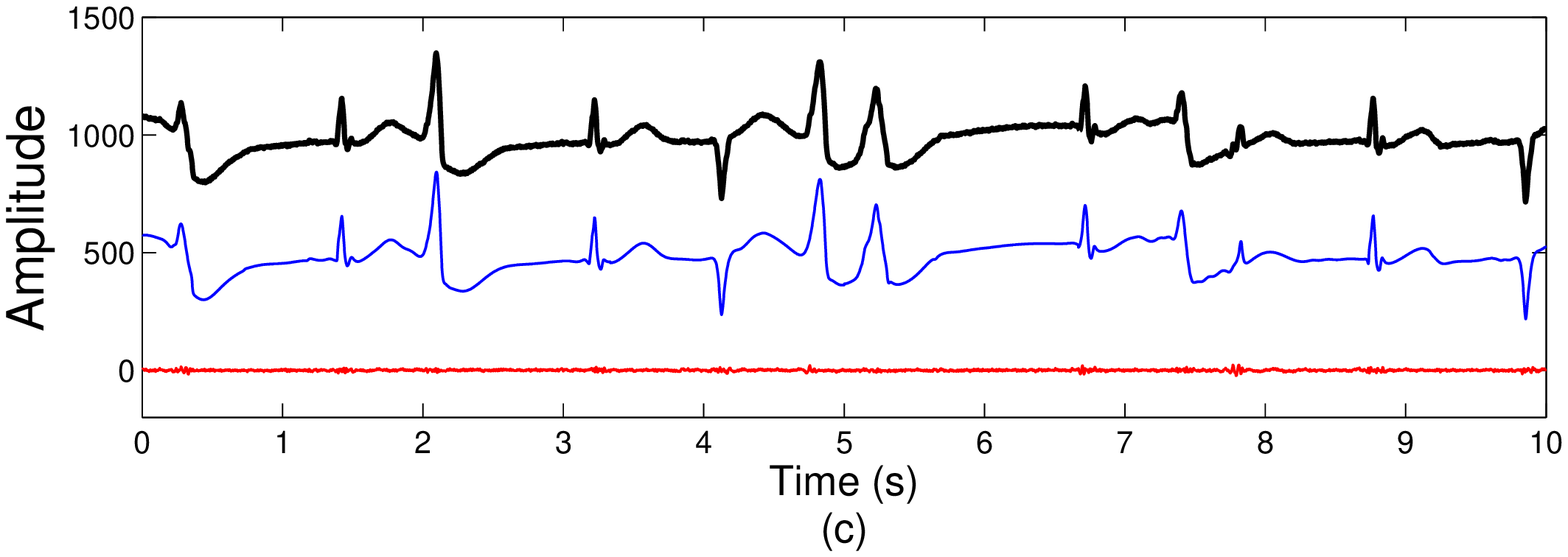}
  \includegraphics[width=.8\linewidth]{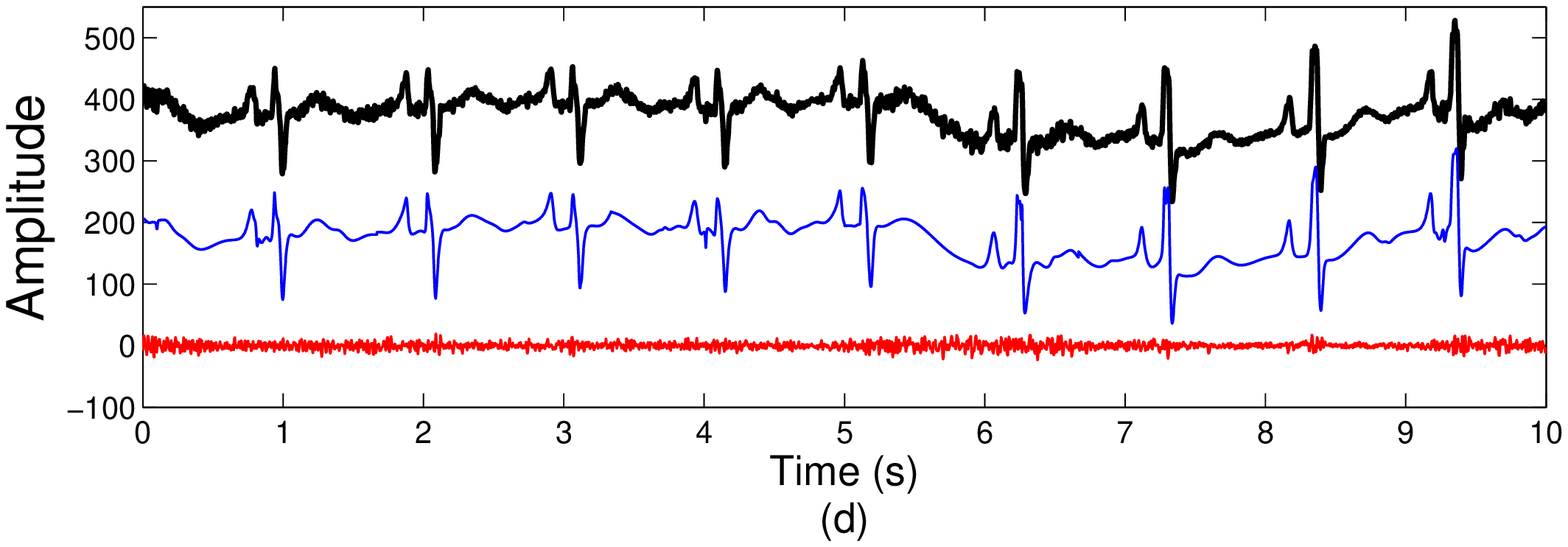}
  \includegraphics[width=.8\linewidth]{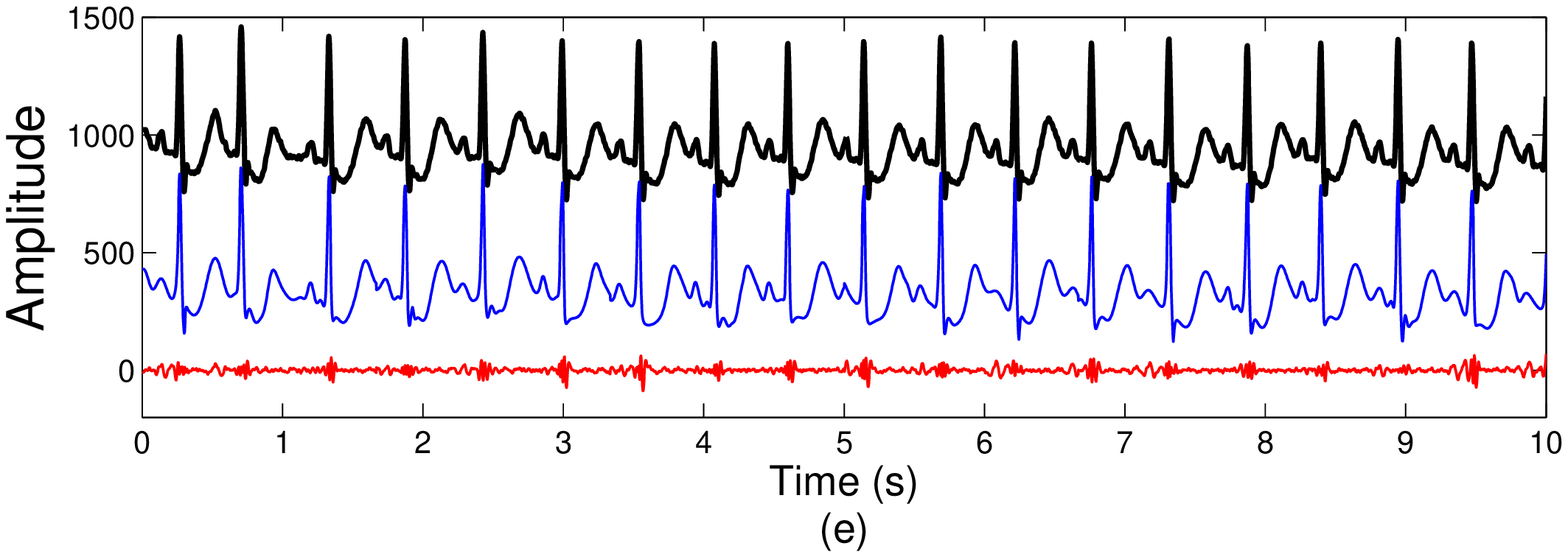}
\caption{Waveforms of several record examples of original (black line), reconstructed (blue line) and error (red line) signals with $N=8$. (a) Record no. 100, CR=39.34, PRD=0.71\%, (b) Record no. 202, CR=33.57, PRD=0.31\%, (c) Record no. 207, CR=35.29, PRD=0.63\%, (d) Record no. 108, CR=35.82, PRD=0.54\%, (e) Record no. 213, CR=34.78, PRD=4.34\%.}
  \label{fig:picture001}
\end{figure}

\subsection{Experimental Results}
The mean of each performance metric over all 48 ECG records of the MIT-BIH Arrhythmia database with different decomposition levels are listed in Table \ref{Table:PerformanceDifferentN}. When $N=7$, the average PRD is $1.71\%$ and the average CR is up to $43.27$. The average CR is $16.85$ and the average PRD decreases to $0.81\%$ when $N$ increases to $14$.
The average CR and PRD of the first compression step, i.e., the performance after applying the Blaschke unwinding AFD compression only, are CR$=23$ and PRD$=1.65\%$ with $N=7$, and CR$=10.74$ and PRD$=0.75\%$ with $N=14$.
The average SNR increases with $N$, which is consistent with PRD. For $N=13$, the average SNR reaches $40$.
QS declines and stabilizes gradually while $N$ increases, which reflects the robustness of the compression efficiency.
Fig.\ref{fig:picture000} shows the mean and standard deviation of four performance metrics over all 48 ECG records for different $N$. Compared with the mean, CR and SNR have small standard deviations. This partially supports that the proposed compression algorithm is applicable to a variety of ECG signals.
The computation time of different $N$ could be found in Table \ref{Table:PerformanceDifferentN}. The compression time is around 1s and increases by 0.1s when $N$ increases by one, while the decomposition time is within dozen of milliseconds.
We remark that in practice, the choice of $N$ depends on the application, since the performance is mainly based on the tradeoff between CR and PRD when $N=7,\ldots,14$. A detailed elaboration will be given in Section \ref{Section:Dicussion}.

For $N=8$, experimental results of all records are given in Table \ref{Table:PerformanceN=8}. Records no. 100 shows a maximum compression rate of $39.34$, and records no. 202 shows a minimum value of $33.57$. Besides, record no. 213 shows the worst PRD of $4.34\%$.
To demonstrate the reconstruction quality, several record examples  for visual inspection are shown in Fig.\ref{fig:picture001}. Record no. 100 represents the waveform and artifact commonly seen in the routine clinical use; Records no. 202, 207 represent complex ventricular, junctional and supraventricular arrhythmias and conduction abnormalities \cite{MM};  Record no. 108 is an ECG signal contains typical noise \cite{MM}, and record no. 213 is with the worst PRD in Table \ref{Table:PerformanceN=8}. From Fig.\ref{fig:picture001} (a), (b), and (c), it is observed that all records are separated deliberately for an easy inspection. Fig.\ref{fig:picture001} (d) validates the proposed method has the ability to denoise signals that is mentioned in the previous section. In Fig.\ref{fig:picture001} (e), the reconstructed signal well preserves the diagnostic features of the original signal despite of poor PRD due to noise smoothing. Altogether, the proposed method can achieve a high quality ECG signal recovery with tiny distortion for a variety of ECG signals.

\begin{figure*}[!htbp]
  \centering
  \includegraphics[width=.99\linewidth]{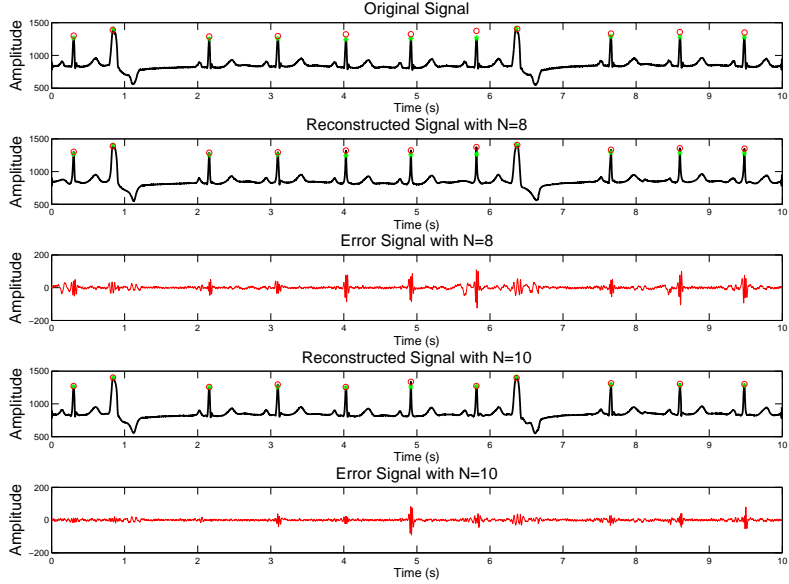}
  \caption{Waveforms of original, reconstructed  and error  signals with $N=8$ and $N=10$ taken from record no.119. The detected QRS complexes of original signal and reconstructed signals are denoted with an asterisk and a circle, respectively.}
  \label{fig:picture002}
\end{figure*}

\subsection{Performance evaluation of QRS detection}

The diagnostic performance of the proposed compression algorithm is assessed by evaluating its ability to preserve the QRS complex.
Since the heart rate information and the heart rate variability (HRV) analysis \cite{TaskForce:1996} is a common target in the e-health field, we focus on how well the R peak information is preserved in the proposed compression algorithm.

We use the state-of-the-art R peak detection algorithm proposed by Elgendi \cite{EM} to detect the QRS complexes. Since there is no preprocessing steps in our proposed compression algorithm, the reconstructed signals can preserve information of the original signal as much as possible. Therefore, the detected R peaks by the Elgendi's algorithm are used as the ground truth to evaluate the accuracy of the QRS complexes of reconstructed signals. A temporal tolerance of $10$ms either side of the annotated QRS complex is deemed as a correct matching tolerance. Note that this is a more stringent criterion compared with other common methods, since generally other authors choose a tolerance greater than $50ms$ \cite{BJJ, CR}.

Fig.\ref{fig:picture002} illustrates the performance of the beat detection algorithm on the original ECG signal and the reconstructed signals with decomposition levels $N=8$ and $N=10$ of Record no. 119 in the MIT-BIH Arrhythmia database. According to the database directory of the MIT-BIH Arrhythmia database \cite{MIT}, Record no. 119 contains normal beats and premature ventricular contraction beats. The signals shown in Fig.\ref{fig:picture002} are over a window of 10s. Based on Fig.\ref{fig:picture002}, we see that all QRS complexes are detected and both reconstructed signal with $N=8$ and $N=10$ preserves important clinical attributes of the original signal. Moreover, the S and T waves are also well preserved that are important for the clinical diagnosis.

\begin{figure}[!htbp]
  \centering
  \includegraphics[width=.6\linewidth]{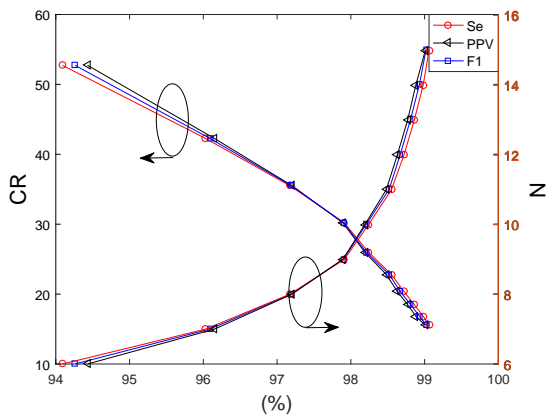}
  \caption{Se, PPV and $F_{1}$ performance in terms of N and CR.}
  \label{fig:picture003}
\end{figure}

The average results of QRS detection performance of all 48 ECG records are displayed in Fig. \ref{fig:picture003}, by showing Se, PPV and $F_{1}$ versus CR and decomposition level $N$. The general trend shown in Fig. \ref{fig:picture003} is that a smaller CR exhibits a better performance in Se, PPV and $F_{1}$. In other words, a higher accuracy in terms of Se, PPV and $F_{1}$ is achieved with a larger $N$. When $N= 6$ and CR equals 50, all Se, PPV and $F_{1}$ values reach to 94\% and approach to 95\%. The high accuracies of Se, PPV and $F_{1}$ reaches 98\% when N is 9 and CR of 30. Furthermore, the performance achieve 99\% accuracy when $N=15$.

\section{Discussions}\label{Section:Dicussion}

\subsection{Performance Comparison}
A comparison of the proposed method and state-of-the-art transform-based ECG compression methods is presented in Table \ref{Table:ComparisonWithOthers}. These methods are tested and validated on the MIT-BIH arrhythmia database. S. Lee et al.\cite{LKL} and A. Pandey et al.\cite{PSSS} are based on DCT,  B. Huang et al.\cite{HWC} and S. Tai et al. \cite{TST} are based on DWT, A. Fathi et al.  \cite{FF} is based on wavelets packets, T. D\'{o}zsa et al. \cite{DK} and R. Kanhe et al. \cite{KH} use Hermite functions, and R. Kumar et al. \cite{KKS}, V. Aggarwal et al. \cite{AP} use SVD and the nonlinear transform method respectively. According to Table \ref{Table:ComparisonWithOthers}, clearly, the average CR of the proposed method is higher than all of the above mentioned methods. Note that we do not compare our results with that reported in \cite{Liu2017}, since only 14 cases out of the whole database were reported in it. We also do not make comparison with the compressed sensing \cite{Zhang2015,Polania2015,Craven2017}, because the algorithm of set partitioning in hierarchical tree (SPIHT) \cite{TST} performs better than the compressed sensing presently, and our performance results outperform SPTHI as shown in Table \ref{Table:ComparisonWithOthers}.
Table \ref{Table:ComparisionWithRecords} lists a comparison for records no. 100, 202, and 207 that represent various characteristics of MIT-BIH arrhythmia signals that are shown in Fig.\ref{fig:picture001} with other methods that provide detailed experimental data. Obviously, for the same record, the proposed method exhibits lower PRD while maintaining similar or even higher CR.

\begin{figure}[!htbp]
  \centering
 \includegraphics[width=.6\linewidth]{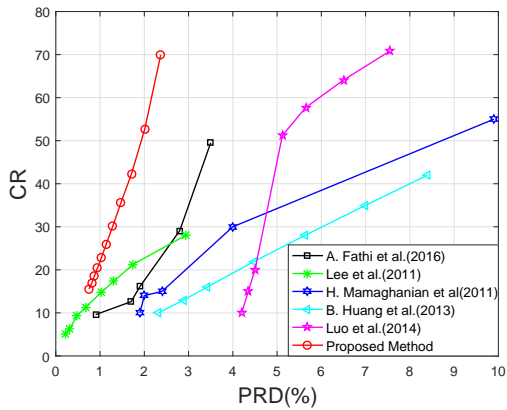}
  \caption{Comparison of the CR of the proposed method in different PRD.}
  \label{fig:picture004}
\end{figure}

Furthermore, Fig.\ref{fig:picture004} plots the PRD-CR relations of the proposed method and some state-of-the-art methods published recently. Clearly, as the CR increases, the PRD increases, i.e., the reconstruction quality decreases. For the same increasing ratio of CR, a smaller increasing ratio of PRD implies a higher QS value. As it can be seen in this figure, the proposed method only has about 0.5\% value added on PRD whenever CR increases by 10, while A. Fathi et al. \cite{FF} and S. Lee et al. \cite{LKL} methods have a near 1\% PRD for CR increases by 10, K. Luo et al. \cite{LLW} has a near  2\% PRD, B. Huang et al. \cite{HWC} method has a more than 2\% PRD and the PRD of H. Mamaghanian et al. \cite{MKV} is more than 4\% for CR increases by 10.
From Fig.\ref{fig:picture004}, for a lower quality, the proposed method also has higher CR and PRD values compared to other methods. When the CR is around 20, the PRD is less than 1\%. When the CR even reaches as high as 30, the PRD is kept less than 2\%.

\begin{figure}[!htbp]
  \centering
 \includegraphics[width=.8\linewidth]{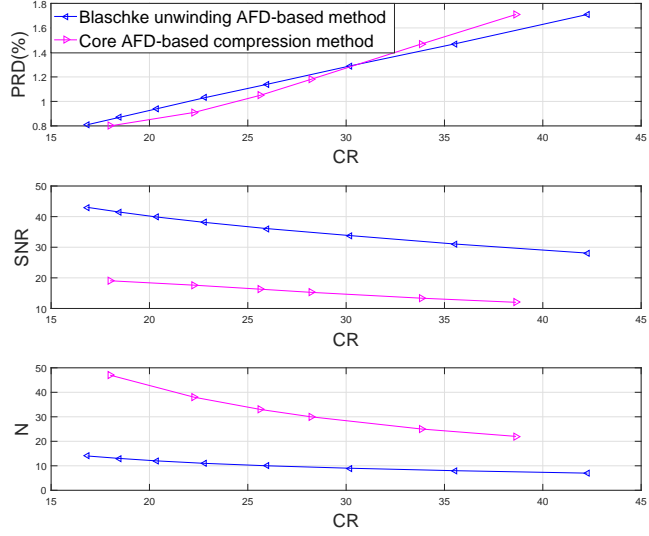}
 \caption{Comparison of the Blaschke unwinding AFD-based compression algorithm with the Core AFD-based compression algorithm.}
  \label{fig:picture005}
\end{figure}
It worths a more detailed comparison of the proposed algorithm and the compression algorithm by using the Core AFD \cite{MZD}. The results of the comparison are shown in Fig. \ref{fig:picture005} in terms of PRD, SNR and N with the different CR. Compared with the Core AFD, the proposed compression algorithm based on the Blaschke unwinding AFD is slightly inferior in the reconstruction quality when CR is less than $30$, while it has a better performance when CR is higher than $30$. Thus the proposed algorithm is more effective in practice since in general we want a higher compression efficiency without lowering the reconstruction quality. Also, the proposed compression algorithm outperforms the Core AFD-based compression algorithm largely in SNR with the same compression efficiency.
Additionally, in our approach, we could obtain a reasonable result when the decomposition level is about $8$ to $10$. However, to have a reasonable result based on the Core AFD, the decomposition level is around $33$. To reduce the parameter numbers, Ma {\em et. al.} in \cite{MZD} modified the Core AFD approach by choosing fixed number of parameters $a_k$ based on the statistical distribution from the randomly chosen $24$ ECG records in the MIT-BIH arrhythmia benchmark database. While this approach leads to a good result, it may run short in the following sense.
Those $24$ ECG signals may not represent all possible characteristics of ECG signals, and an overfitting might happen. Also, the fixed parameter $a_k$ weakened the adaptivity spirit of the AFD. Since the parameters are determined from the ECG signals, it may not be applied to other signals.
Furthermore, Ma et al in \cite{MZD} only pass the sequence number of each parameter $a_k$ instead of passing all the compression data to the user during the data transmission. As a result, they need to send all the fixed sorted parameter $a_k$ sequence to users in advance. During the uncompressing process, users need to find the corresponding parameter $a_k$ by its sequence number. This situation restricts the application of the algorithm. Finally, Ma et al \cite{MZD} did not assess how their algorithm influences the R peak detection algorithm.

To the best of our knowledge, previous works in the field mainly focused on the signal fidelity such as PRD, but not the signal quality for the diagnostic applications like the R peak detection, So there is no comparison available.


\begin{table}[!htbp]
\centering\caption{Comparison between the average performance of the proposed method and other methods\label{Table:ComparisonWithOthers}}
\begin{tabular}{ccccccc}
\toprule
Method            & CR     & PRD     & QS       \\\midrule

B. Huang et al.\cite{HWC}           & 13.00 & 2.89 & --   \\
T. D\'{o}zsa et al. \cite{DK} &14.25 & 22.21 & 1.77\\
R. Kanhe et al. \cite{KH}     &3.44 & 15.31 &--\\
S. Lee et al.\cite{LKL}          &5.19  &\textbf{0.23}  & 22.32  \\
R. Kumar et al. \cite{KKS}   & 21.91  & 1.93 &--\\
J. Ma et al. \cite{MZD}       &25.64  & 1.05  & 30.51 \\
A. Fathi et al.  \cite{FF}          &29.10  &2.83  &-- \\
S. Tai et al. \cite{TST}            & 8.00      & 1.18   & 6.78  \\
V. Aggarwal et al. \cite{AP}    &28.22  &2.65  & --   \\
A. Pandey et al.\cite{PSSS} &18.59 & 1.06 &17.57\\\bottomrule
Proposed (N=8)           &\textbf{35.53}  &1.47  &\textbf{32.58}  \\\bottomrule
\end{tabular}
\end{table}

\begin{table}[!htbp]
\centering\caption{Performance comparison of some ECG compression schemes for some different records\label{Table:ComparisionWithRecords}}
\begin{tabular}{cccc|ccc|ccc}
\toprule
Algorithm                       & Record    & CR      & PRD   & Record    & CR      & PRD   & Record    & CR      & PRD\\\midrule
Lee  et al.\cite{LKL}           &100        &22.94    &1.95   &202        &21.66    &0.91   &207        &24.45    &0.61\\
A. Fathi et al.  \cite{FF}      &100        &31.42    &3.21   &202        &26.5     &2.9    &207        &29.4     &2.1\\
J. Ma et al. \cite{MZD}         &100        &25.64    &0.57   &202        &25.64    &0.33   &207        &25.64    &0.51\\
A. Pandey et al.\cite{PSSS}     &100        &16.87    &1.00   &202        &26.34    &1.01   &207        &31.12    &14.85\\\midrule
Proposed(N=8)                   &100        &39.34    & 0.71  &202        &33.57    &0.31   &207        &35.29    &0.63\\
Proposed(N=10)                  &100        &26.09    & 0.57  &202        &26.09    &0.30   &207        &26.52    &0.27\\\bottomrule

\end{tabular}
\end{table}

\subsection{Setting the value of $N$}
$N$ is the key parameter of the proposed compression algorithm, and it balances the compression efficiency and the reconstruction quality, as is shown in Table \ref{Table:PerformanceDifferentN}. When the decomposition level is $N$, there are $2N+1$ parameters and several zeros that need to be saved for the compression, and the reminder is $R_{N}$. When $N$ increases, the number of parameters and zeros increases and $R_{N}$ gets smaller, and these are directly reflected in CR and PRD respectively; that is, the compression efficiency declines while the reconstruction quality becomes better as a compromise. The choice of $N$ depends on the application. In the practical e-health applications, after setting an initial value of $N$, $N$ can be adjusted to guarantee the requirement of the compression efficiency and the reconstruction quality. Specifically, depending on the scenario, designers or users can lower $N$ to obtain a high compression efficiency under a specified fidelity, or increase $N$ to achieve a better reconstruction quality in exchange of a lower compression efficiency.

\subsection{Other practical issues}

A separate but relevant topic that we do not discuss in this paper is the data encryption issue, which is needed to protect the patient privacy issue when any bio-signal is transmitted over public networks, according to the Health Insurance Portability and Accountability Act (HIPAA) \cite{Liu2017, HIPPA}. A study taking into account the encryption to the proposed compression algorithm will be reported in the future work. As the technology advances, more channels could be recorded and transmitted. An immediate question would be if we could simultaneously compress those channels. In \cite{Capurro2017}, a sequential compression algorithm for multichannel biomedical signals is proposed to handle this challenge. We will generalize the proposed unwinding AFD framework to the multichannel case in the future work, and develop the associated mathematical framework.

Another important topic associated with the e-health is the real-time implementation of the proposed compression algorithm. As is shown in Table \ref{Table:PerformanceDifferentN}, the compression time and the decomposition time take in total about 1 second and is relatively short compared with the length of the signal under compression, which is $600/360\approx 2$ seconds. Therefore, the proposing method could be applied to carry out the real-time compression and transmission.

\subsection{Limitation and future work}
Despite the strength of the proposed algorithm, we acknowledge several limitations of the current study. First, the data is from a publicly available database that is not collected from an equipment tailored for the telemedicine. Also, the database size is limited. Therefore, a larger database collected from such an equipment is needed to confirm the performance of the proposed algorithm. Second, we focus only on the ECG signal in this paper, while the proposed algorithm has a potential to be applied to other biomedical signals. We will explore this possibility in the future work. Third, we take the output of the Elgendi's R peak detection algorithm as the truth to evaluate how well the R peak is preserved by the proposed compression algorithm. We could therefore only conclude that the compression algorithm works well with the Elgendi's R peak detection algorithm. However, there are several R peak detection algorithms based on different philosophy, and it is not clear if the proposed compression algorithm works well with them. Furthermore, we only focus on the R peak detection, while there are other clinically important landmarks in the ECG signal that we do not explore. A systematic study for this issue will be reported in the future work.

\section{Conclusion}\label{Section:Conclusion}
In this paper, a novel ECG compression method based on the Blaschke unwinding AFD and the Huffman encoding is presented,
which leads to a high compression rate with a high fidelity. Compared with existing algorithms, like wavelet transform, DCT, CS, SVD, etc, the proposed algorithm is free of any preprocessing of the ECG signal and efficient evaluated from different aspects.
The decomposition level $N$ and the PRD with different CR values are investigated with 48 ECG records from MIT-BIH arrhythmia benchmark database. We further show that the ECG morphological characteristic is well preserved in the reconstructed signals by evaluating how a R peak detection algorithm is influenced.
This method has a potential to be used as a compression method of ECG signals for telemedicine and storage of ECG signals in e-health.
Source codes of the paper has been put on the Code Ocean website, which can be found in the following address: https://codeocean.com/2018/01/25/blaschke-unwinding-afd-based-ecg-compression/code.

\section*{Acknowledgment}

This study is supported by the research grant: MYRG2014-00009-FST, YRG2017-00218-FST and The Science and Technology Development Fund of Macao SAR FDCT 079/2016/A2.


\end{document}